\documentclass[prb,twocolumn,showpacs,superscriptaddress,preprintnumbers,floatfix,amsmath,amssymb]{revtex4-1}
\usepackage{amssymb}
\usepackage{graphicx}% Include figure files
\usepackage{dcolumn}% Align table columns on decimal point
\usepackage{bm}% bold math
\usepackage{color}
\hfuzz=\maxdimen
\begin{document}

\title{Unconventional superconductivity in quasi-one-dimensional Rb$_2$Cr$_3$As$_3$ }

\author{Zhang-Tu Tang}
\affiliation{Department of Physics, Zhejiang University, Hangzhou
310027, China}

\author{Jin-Ke Bao}
\affiliation{Department of Physics, Zhejiang University, Hangzhou
310027, China}

\author{Yi Liu}
\affiliation{Department of Physics, Zhejiang University, Hangzhou
310027, China}

\author{Yun-Lei Sun}
\affiliation{Department of Physics, Zhejiang University, Hangzhou
310027, China}

\author{Abduweli Ablimit}
\affiliation{Department of Physics, Zhejiang University, Hangzhou
310027, China}

\author{Hui-Fei Zhai}
\affiliation{Department of Physics, Zhejiang University, Hangzhou
310027, China}

\author{Hao Jiang}
\affiliation{Department of Physics, Zhejiang University, Hangzhou
310027, China}

\author{Chun-Mu Feng}
\affiliation{Department of Physics, Zhejiang University, Hangzhou
310027, China}

\author{Zhu-An Xu}
\affiliation{Department of Physics, Zhejiang University, Hangzhou
310027, China} \affiliation{State Key Lab of Silicon Materials,
Zhejiang University, Hangzhou 310027, China}
\affiliation{Collaborative Innovation Centre of Advanced Microstructures, Nanjing 210093, China}

\author{Guang-Han Cao} \email[Correspondence should be sent to: ]{ghcao@zju.edu.cn}
\affiliation{Department of Physics, Zhejiang University, Hangzhou
310027, China} \affiliation{State Key Lab of Silicon Materials,
Zhejiang University, Hangzhou 310027, China} \affiliation{Collaborative Innovation Centre of Advanced Microstructures, Nanjing 210093, China}

\date{\today}

\begin{abstract}
Following the discovery of superconductivity in quasi-one-dimensional K$_2$Cr$_3$As$_3$ containing [(Cr$_3$As$_3$)$^{2-}$]$_{\infty}$ chains [J. K. Bao et al., arXiv: 1412.0067 (2014)], we succeeded in synthesizing an analogous compound, Rb$_2$Cr$_3$As$_3$, which also crystallizes in a hexagonal lattice. The replacement of K by Rb results in an expansion of $a$ axis by 3\%, indicating a weaker interchain coupling in Rb$_2$Cr$_3$As$_3$. Bulk superconductivity emerges at 4.8 K, above which the normal-state resistivity shows a linear temperature dependence up to 35 K. The estimated upper critical field at zero temperature exceeds the Pauli paramagnetic limit by a factor of two. Furthermore, the electronic specific-heat coefficient extrapolated to zero temperature in the mixed state increases with $\sqrt{H}$, suggesting existence of nodes in the superconducting energy gap. Hence Rb$_2$Cr$_3$As$_3$ manifests itself as another example of unconventional superconductor in the Cr$_3$As$_3$-chain based system.
\end{abstract}

%74.70.-b    Superconducting materials other than cuprates
%74.70.Dd Ternary, quaternary, and multinary compounds (including Chevrel phases, borocarbides, etc.)
%71.45.Lr    Charge-density-wave systems (see also 75.30.Fv Spin-density waves)
%74.62.Bf    Effects of material synthesis, crystal structure, and chemical composition (for methods of materials synthesis, see 81.20.-n)
%61.66.Fn Inorganic compounds
%74.25.Bt Thermodynamic properties
\maketitle

Superconductivity (SC) infrequently appears in a quasi-one-dimensional (Q1D) crystalline material primarily because of Peierls instability towards a state of charge-density wave.\cite{smaalen,gruner} Additional disadvantage of Q1D systems for SC may come from the possible realization of Luttinger liquid, as well as the reduced density of states when the Fermi level ($E_{\text{F}}$) does not incidentally locate at the Van Hove singularities. Nevertheless, 1D systems have long been studied theoretically\cite{voit} owing to their inherent simplicity and, the discovery of SC in Q1D Bechgaard salts\cite{jerome} has attracted sustained research interest.\cite{wzhang} Examples of inorganic Q1D superconductors include Li$_{0.9}$Mo$_6$O$_{17}$\cite{greenblatt} and Tl$_2$Mo$_6$Se$_6$\cite{armici}, in which the Mo-4$d$ electrons bear relatively weak electron correlations. Nevertheless, peculiar properties were revealed in Li$_{0.9}$Mo$_6$O$_{17}$,\cite{denlinger,xxf,Hc2_LiMoO} and possible triplet SC was proposed.\cite{lebed}

Very recently, we discovered bulk SC at 6.1 K in a Q1D pnictide K$_2$Cr$_3$As$_3$ in which correlated Cr-3$d$ electrons are involved.\cite{bao} The crystal structure is characterized by the [(Cr$_3$As$_3$)$^{2-}$]$_{\infty}$ double-walled subnano-tubes in which Cr$_{6/2}$ (As$_{6/2}$) face-sharing octahedron chains constitute the inner (outer) wall. The intrachain Cr$-$Cr bond distances (2.6 $\sim$ 2.7 $\mathrm{\AA}$) are close to that in Cr metal, indicating a dominant metallic bonding along the chain direction. In comparison, the shortest interchain Cr$-$Cr distance is as large as 7.3 $\mathrm{\AA}$, and a much weaker (if not negligible) interchain overlap of Cr-3$d$ orbitals is expected. Thus, the material is really Q1D at least in the sense of chemical bonding as well as crystal structure. The new material hosts series of peculiar physical properties including a large Sommerfeld coefficient of 70 $\sim$ 75 mJ K$^{-2}$ mol$^{-1}$, a linear temperature dependence of resistivity from 7 to 300 K, and high upper critical fields exceeding the Pauli limit by a factor over three. The first-principles calculations\cite{jh} indicate that the electrons around $E_{\text{F}}$ are dominated by the Cr-3$d$ states, which form two Q1D Fermi surface (FS) sheets and one 3D FS pockets. The calculated 'bare' density of state (DOS) at $E_{\text{F}}$ is less than 1/3 of the experimental value from the specific-heat measurement, confirming the electron correlation effect. Besides, ferromagnetic instability and/or ferromagnetic spin fluctuations are revealed, which favors spin-triplet Cooper pairing. These experimental and theoretical-calculation results suggest unconventional SC in K$_2$Cr$_3$As$_3$.

In Q1D systems, the interchain coupling plays a crucial role in the competition between SC and charge- or spin-density wave. For example, unlike Tl$_2$Mo$_6$Se$_6$ which remains metallic and superconducts at 4.2 K,\cite{petrovic} Rb$_2$Mo$_6$Se$_6$ undergoes a metal-insulator transition at $\sim$ 170 K, which is explained in terms of dynamical charge-density wave owing to the reduced interchain hopping.\cite{petrovic} Thus, the appearance of SC in K$_2$Cr$_3$As$_3$ suggests a significant interchain coupling, as revealed by the existence of 3D FS. Therefore, it is of considerable interest to investigate the effect of interchain coupling by the elemental replacement (e.g., Rb for K) in the spacer columns. Furthermore, it is important to check whether the unconventional SC is a common feature in the Q1D Cr-based pnictides.

In this Communication, we report the synthesis, structural characterization, and physical property measurement of an analogous material Rb$_2$Cr$_3$As$_3$. Indeed, a weaker interchain coupling is signatured by the obviously expanded interchain distance. Nevertheless, bulk superconductivity at a little lowered temperature of 4.8 K was observed, accompanied with a mildly reduced Sommerfeld coefficient of 55 mJ K$^{-2}$ mol$^{-1}$. Remarkably, unconventional SC is evidenced by the electrical transport and the thermodynamic properties.

The Rb$_2$Cr$_3$As$_3$ polycrystalline sample was synthesized by solid state reactions in sealed vacuum. First, the starting materials (Rb piece, Cr powder and As piece) all with high purity ($\geq$99.9\%, Alfa Aesar) were weighed in the nearly stoichiometric ratio (with 3\% excess of Rb in order to compensate the loss of Rb in the process of the solid-state reaction), and the mixture was loaded in a quartz ampoule. The quartz ampoule was evacuated ($<10^{-2}$Pa) and sealed, followed by slowly heating in a muffle furnace to 523 K for 15 h. After the first stage reaction, the mixture was homogenized by grinding, pressed into pellets, and put into an alumina tube which was then sealed by arc welding in argon atmosphere in a Ta tube. The sealed Ta tube, jacketed by an evacuated quartz ampoule, was then sintered at 973 K for 24 h. This procedure was repeated to allow a full solid-state reaction. The obtained sample was sensitive to air, and it should not be exposed in air as far as possible. Note that all the operations of weighing, mixing, grinding, pelletizing, etc., were carried out in an argon-filled glovebox with the water and oxygen contents less than 0.1 ppm.

Powder x-ray diffraction (XRD) was carried out at room temperature on a PANalytical x-ray diffractometer (Model EMPYREAN) with a monochromatic CuK$_{\alpha1}$ radiation. To avoid exposure to air, the sample powder was mixed with small amount of Apiezon $N$ grease before putting on the sample holder, and the sample was in vacuum during the XRD data collecting. The crystal structure parameters were refined based on the K$_2$Cr$_3$As$_3$-type structure\cite{bao} by a Rietveld analysis with the code RIETAN-FP.\cite{rietan-fp} Note that the XRD data of 2$\theta$ = 8$^{\circ}$ $\sim$ 25$^{\circ}$ were not included for the refinement, to avoid the extrinsic influence from the $N$ grease.

We conducted the electrical, magnetic and heat capacity measurements on a physical property measurement system (Quantum Design, PPMS-9) and a magnetic property measurement system (Quantum Design, MPMS-5). The electrical resistivity was measured using a standard four-electrode method. The as-prepared Rb$_2$Cr$_3$As$_3$ pellet was cut into a thin rectangular bar, and gold wires were attached onto the sample's surface with silver paint. The heat capacity was measured by a thermal relaxation method using a square-shaped sample plate (9.0 mg). The temperature dependence of dc magnetic susceptibility was measured in the protocols of zero-field cooling (ZFC) and field cooling (FC) under a field of 10 Oe. The dc magnetization was also measured as a function of magnetic field at 2 K.

\begin{figure}
\includegraphics{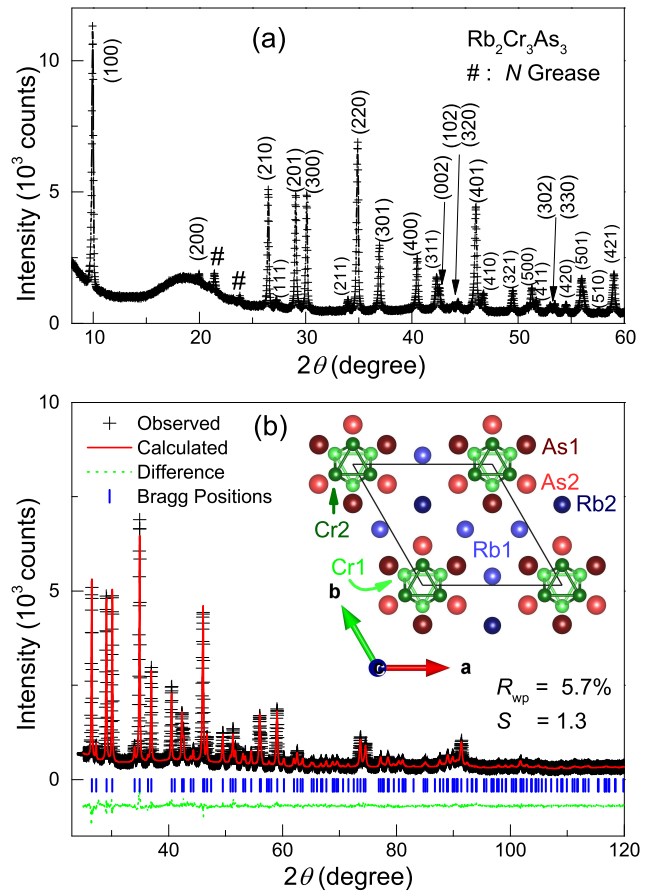}
\caption{\label{XRD}(Color online) (a) Powder X-ray diffraction of the Rb$_2$Cr$_3$As$_3$ polycrystalline sample protected by the Apiezon $N$ grease, which is indexed by a hexagonal unit cell. (b) Rietveld refinement profile for the powder x-ray diffraction with 25$^{\circ}\leq 2\theta \leq 120^{\circ}$). $R_{\mathrm{wp}}$ and $S$ denotes the weighted profile reliable factor and "goodness-of-fit" parameter, respectively. The inset shows the crystal structure projected along the \textbf{c} axis.}
\end{figure}

Figure~\ref{XRD} shows the XRD profile of the Rb$_2$Cr$_3$As$_3$ sample. The reflections can be well indexed by a hexagonal unit cell whose size is slightly larger than that of K$_2$Cr$_3$As$_3$, and no obvious impurity phase can be identified. The Rietveld refinement was successful, and the resulted structural parameters are listed in Table ~\ref{parameter}. Compared with K$_2$Cr$_3$As$_3$, the $a$ axis is 3.0\% larger, while the $c$ axis is expanded merely by 0.28\%. Since the $a$ axis just measures the interchain distance [see the inset of Fig.~\ref{XRD}(b)], one may expect a weaker interchain interaction in Rb$_2$Cr$_3$As$_3$. This is further revealed by the enlargement of the interchain Cr$-$Cr distance (from 7.30 $\mathrm{\AA}$ in K$_2$Cr$_3$As$_3$ to 7.63 $\mathrm{\AA}$ in Rb$_2$Cr$_3$As$_3$), while the average Cr$-$Cr bond distance almost keeps unchanged. Besides, the [(Cr$_3$As$_3$)$^{2-}$]$_{\infty}$ column itself suffers significant modifications. The Cr2$-$Cr2 bond distance of the Cr2 triangle in the $z$ = 0 plane \emph{decreases} by 4.5\%, and the Cr1$-$Cr1 bond distance in the $z$ =0.5 plane \emph{expands} by 1.5\%, as compared with those in K$_2$Cr$_3$As$_3$. Consequently, the geometry of the Cr octahedron becomes inverse.

\begin{table}
\caption{\label{parameter}
Comparison of crystallographic data and some physical parameters of $A_2$Cr$_3$As$_3$ [$A$ = K\cite{bao} and Rb (present work)]. The atomic coordinates are as follows: As1 ($x$, $-x$, 0); As2 ($x$, $-x$, 0.5); Cr1 ($x$, $-x$, 0.5); Cr2 ($x$, $-x$, 0); $A$1 ($x$, $-x$, 0.5); $A$2 (1/3, 2/3, 0). $T_{\text{c}}$, $\Delta T_{\text{c}}$, $H_{c2}$, $\gamma_\mathrm{n}$, $\theta_{\text{D}}$, and $\Delta C$ denote the superconducting critical temperature, transition width, upper critical field, electronic specific-heat coefficient, Debye temperature, and dimensionless specific-heat jump, respectively.}
\begin{ruledtabular}
\begin{tabular}{ccccc}
  & K$_2$Cr$_3$As$_3$&  &Rb$_2$Cr$_3$As$_3$& \\
\hline
Space Group&   $P \overline{6}$ $m$ 2 & & $P \overline{6}$ $m$ 2  \\\hline
$a$ (\r{A})  & 9.9832(9)&  &10.281(1)& \\
$c$ (\r{A})  & 4.2304(4)&  &4.2421(3)& \\
$V$ (\r{A}$^{3}$)   & 365.13(6)&  &388.32(5)&  \\
\hline
Coordinates ($x$)& &  &  & \\ \hline
As1 (3$j$)&0.8339(2)&   &0.8382(2)  &\\
As2 (3$k$)&0.6649(4)&  &0.6727(2)  & \\
Cr1 (3$k$)&0.9127(3)&  &0.9140(2)  &\\
Cr2 (3$j$)&0.8203(6)&   &0.8333(3)  &\\
$A$1 (3$k$)&0.5387(4)&   &0.5372(1)  & \\
$A$2 (1$c$)&0.3333&   &0.3333 \\
\hline
Bond Distances && &&  \\\hline
Cr1$-$Cr1 (\r{A}) & 2.614(9) &   & 2.654(4) &  \\
Cr1$-$Cr2 (\r{A}) & 2.612(2) &   & 2.603(2) &  \\
Cr2$-$Cr2 (\r{A}) & 2.69(1) &   & 2.570(3) &  \\\hline
Physical Property Parameters && &&  \\\hline
$T_{\text{c}}$ (K)  & 6.1 &   & 4.8 &  \\
$\Delta T_{\text{c}}$ (K)  & 0.24 &   & 0.44 &  \\
$\mu_{0}$(d$H_{c2}$/d$T$)$_{T_{c}}$ (T/K) & $-$7.43 &  & $-$5.08 & \\
$\gamma_\mathrm{n}$ (mJ/K$^2$/mol) & 70 $\sim$ 75 &   & 55 &  \\
$\theta_{\text{D}}$ (K) & 215 $\sim$ 218 &   & 175 &  \\
$\Delta C/(\gamma_{\text{n}} T_{\text{c}})$ & 2.0 $\sim$ 2.4 &   & 1.8 &  \\
\end{tabular}
\end{ruledtabular}
\end{table}

\begin{figure}
\includegraphics{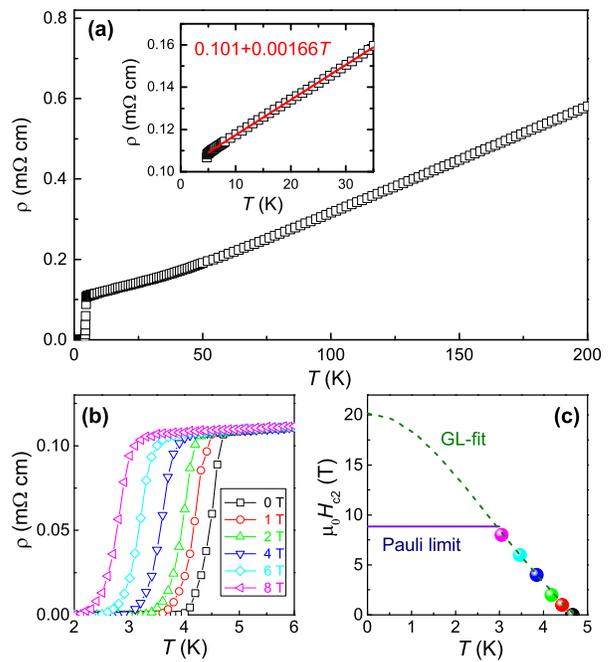}
\caption{\label{resistivity} (Color online) (a) Temperature dependence of the electrical resistivity for Rb$_2$Cr$_3$As$_3$. The inset show the linear temperature dependence from 5 to 35 K. (b) Superconducting transitions under magnetic fields up to 8 T. (c) The upper critical field as a function of temperature. The dashed line denotes a Ginzburg-Landau (GL) fit. The Pauli-limit field ($\mu_{0}H_{\text{P}} = 1.84 T_{\text{c}}$) is labelled by the horizontal line.}
\end{figure}

Figure~\ref{resistivity}(a) shows temperature dependence of resistivity, $\rho(T)$, for the Rb$_2$Cr$_3$As$_3$ polycrystalline sample. The $\rho(T)$ shows metallic conduction and, similar to that in K$_2$Cr$_3$As$_3$, a $T$-linear resistivity can also be seen, albeit in a narrower temperature range from 5 to 35 K. This indicates a non-Fermi-liquid (NFL) behavior, if it is not an accidental phenomenon due to a "polycrystal effect".\cite{polycrystal} The origin of the NFL behavior could be related to a Tomonaga-Luttinger liquid state,\cite{NMR} characteristic of 1D fermion system.\cite{voit,ogata-anderson} Nonetheless, superconductivity emerges below 4.8 K with a transition width $\Delta T_{\text{c}}$ = 0.44 K [$\Delta T_{\text{c}} = T(90\%\rho_{\text{n}}) - T(10\%\rho_{\text{n}})$, where $\rho_{\text{n}}$ refers to the extrapolated normal-state resistivity at the superconducting transition midpoint]. This $\Delta T_{\text{c}}$ value is obviously larger than that (0.24 K) of K$_2$Cr$_3$As$_3$, which seems to be related to the relatively weak interchain coupling in Rb$_2$Cr$_3$As$_3$. Other explanations of the broadening of superconducting transition include less inhomogeneity and/or larger stresses/strains in the present polycrystalline sample.

Upon applying magnetic field, $T_{\text{c}}$ decreases monotonically, as shown in Fig.~\ref{resistivity}(b), from which the upper critical field, $H_{\text{c2}}$, can be determined using the conventional criteria of $90\%\rho_{\text{n}}$. The resulted $H_{\text{c2}}$ is plotted as a function of temperature in Fig.~\ref{resistivity}(c). The initial slope $\mu_{0}$(d$H_{c2}$/d$T$)$_{T_{c}}$ is $-$5.08 T/K. If using the empirical Ginzburg-Landau relation,\cite{GL} $H_{\text{c2}}(T) = H_{\text{c2}}(0)(1-t^{2})/(1+t^{2})$, where $t$ denotes a reduced temperature $T/T_{\text{c}}$, the upper critical field at zero temperature [$\mu_{0}H_{\text{c2}}(0)$] is estimated to be $\sim$ 20 T. Alternatively, the orbital limited upper critical field $\mu_{0}H_{\text{c2}}^{\mathrm{orb}}(0)$ can be estimated by the Werthammer-Helfand-Hohenberg model,\cite{WHH,hake} which yields $\mu_{0}H_{\text{c2}}^{\mathrm{orb}}(0) = 0.693T_{c}$(d$\mu_{0}H_{c2}$/d$T$)$_{T_{c}}\approx$ 17 T for an isotropic full superconducting gap in the dirty limit and without spin-orbit coupling. Both the $\mu_{0}H_{\text{c2}}(0)$ values exceed the Pauli paramagnetic limited upper critical field of $\mu_{0}H_{\text{\text{P}}}$ = 1.84 $T_{\text{c}}$ $\approx$ 9.0 T (also for an isotropic full superconducting gap without spin-orbit coupling).\cite{clogston,chandrasekhar} Note that the $H_{\text{c2}}(T)$ data were measured on the polycrystalline sample, and one may expect a higher $H_{c2}(0)$ for $\textbf{H}\|\textbf{c}$. A similar phenomenon of large $H_{c2}(0)$ beyond the Pauli limit was reported in Q1D  Li$_{0.9}$Mo$_6$O$_{17}$,\cite{Hc2_LiMoO} which was regarded as an evidence of spin-triplet SC.\cite{lebed} For the Q1D superconductor Tl$_2$Mo$_6$Se$_6$ ($T_\mathrm{c}$ = 4.2 K), in comparison, the measured $\mu_{0}H_{c2}(0)$ values are 5.8 T and 0.47 T, respectively, for magnetic fields parallel and perpendicular to the $c$ axis,\cite{petrovic} both of which are less than the Pauli limited upper critical field.

\begin{figure}
\includegraphics{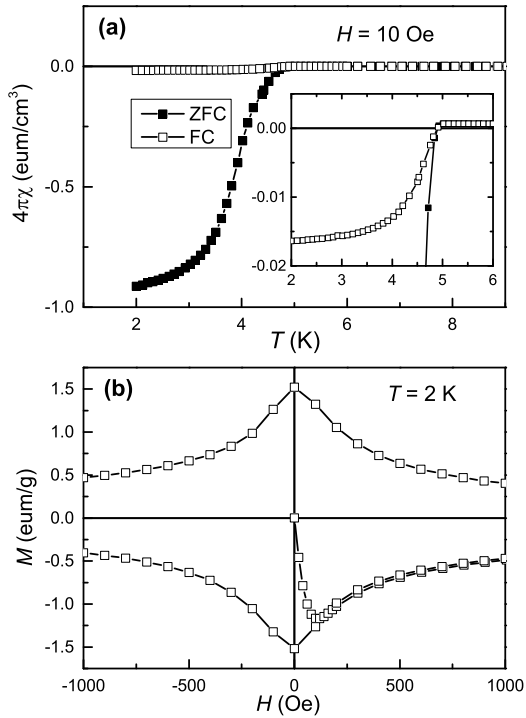}
\caption{(a) Temperature dependence of magnetic susceptibility (scaled by 4$\pi\chi$) for the Rb$_2$Cr$_3$As$_3$ polycrystals. The inset zooms in the superconducting transition in field-cooling (FC) mode. (b) Field dependence of magnetization at 2 K. The magnetic field was applied in the order of $0\rightarrow$ 5 T$\rightarrow -5$ T$ \rightarrow $ 5 T.\label{magnetization}}
\end{figure}

In order to verify whether the observed SC is of bulk nature, we performed dc magnetic susceptibility measurements in both ZFC and FC modes under a 10 Oe magnetic field. Indeed, a superconducting diamagnetic transition can be seen at $T_{\text{c}}$ = 4.8 K in Fig.~\ref{magnetization}(a). The volume fraction of magnetic shielding scaled by 4$\pi\chi_{\text{ZFC}}$ is 91\% at 2 K after the correction of demagnetization effect, suggesting bulk SC. In contrast, the volume fraction of magnetic repulsion (4$\pi\chi_{\text{FC}}$) is merely 1.6\%. This could be due to the magnetic flux pinning in the process of cooling down under magnetic fields. It is also noted that the diamagnetic transition width is significantly broadened, compared with K$_2$Cr$_3$As$_3$\cite{bao}, consistent with the above resistivity measurement result. Fig.~\ref{magnetization}(b) shows the field dependence of magnetization at 2 K, indicating an extreme type-II SC with very small lower critical field (below 40 Oe). The magnetic flux pinning effect is evident from the magnetic hysteresis loop.

\begin{figure}
\includegraphics{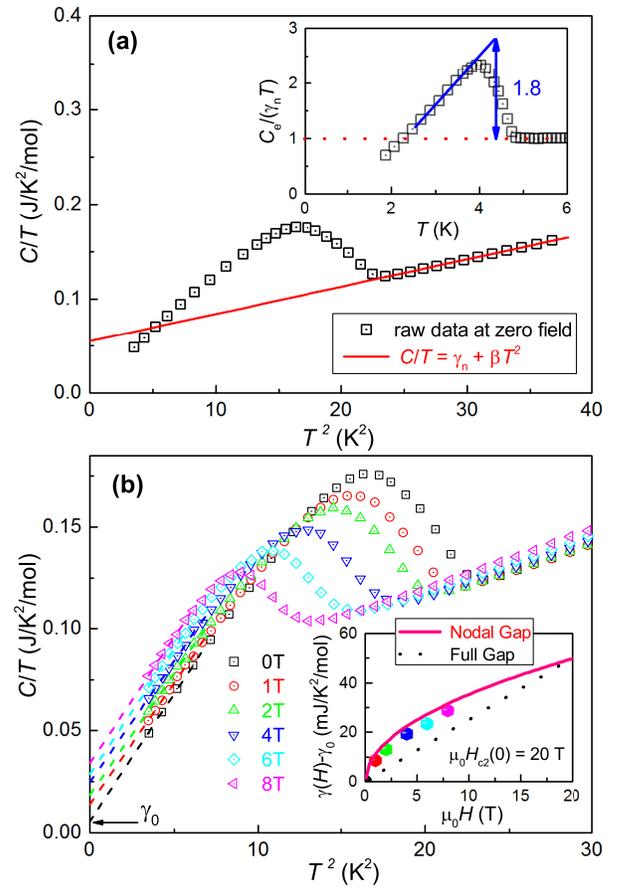}
\caption{\label{specific-heat}(Color online) Low-temperature specific-heat data for Rb$_2$Cr$_3$As$_3$, plotted with $C/T$ vs $T^2$. The inset of (a) shows the normalized electronic specific heat, $C_{\text{e}}(T)/(\gamma_\mathrm{n} T)$, as a function of temperature. Panel (b) presents the $C(T)$ data under magnetic fields, from which the zero-temperature Sommerfeld coefficients in the superconducting mixed state, $\gamma(H)$ can be extrapolated. The inset displays the $\gamma(H)$ dependence, in which theoretical prediction for a nodal (full) superconducting gap is shown by the solid (dotted) line. }
\end{figure}

Figure~\ref{specific-heat} shows temperature dependence of specific heat [$C(T)$] in the low-$T$ regime, plotted in a $C/T$ vs $T^2$ scale. In a short temperature range from 4.8 to 6 K, the $C/T$ vs $T^2$ is essentially linear, and the linear fit with a formula $C/T = \gamma_{\text{n}} + \beta T^2$ yields an intersect of $\gamma_{\text{n}}$ = 55.1 mJ K$^{-2}$ mol$^{-1}$ and a slope of $\beta$ = 2.9 mJ K$^{-4}$ mol$^{-1}$. The $\gamma_{\text{n}}$ value refers to as the normal-state Sommerfeld coefficient, which is lower but comparable to that of K$_2$Cr$_3$As$_3$ (70 $\sim$ 75 mJ K$^{-2}$ mol$^{-1}$).\cite{bao} With the fitted $\beta$ value, and using the formula $\theta_{\text{D}}=[(12/5)NR\pi^{4}/\beta]^{1/3}$, the Debye temperature can be calculated to be 175 K, which is reasonably smaller than that in K$_2$Cr$_3$As$_3$ (215 $\sim$ 218 K).

Bulk SC is further confirmed by the characteristic specific-heat jump [$\Delta(C)$] below 4.8 K. In general, the phonon contribution to the specific heat in the superconducting state follows the relation fitted from the normal-state $C(T)$ data. Hence the electronic specific heat [$C_{\text{e}}(T)$] can be obtained by the deduction of $\beta T^{3}$ from the total $C(T)$.~\cite{Ce} The inset of Fig.~\ref{specific-heat}(a) shows the obtained $C_{\text{e}}(T)$ (normalized by $\gamma_{\text{n}} T$) as a function of temperature. By the entropy conserving construction, the thermodynamic transition temperature is determined to be 4.4 K. The dimensionless specific-heat jump at $T_{\text{c}}$ [$\Delta C/(\gamma_{\text{n}} T_{\text{c}})$] is about 1.8, substantially lower than that (2.0 $\sim$ 2.4) in K$_2$Cr$_3$As$_3$, which seems to be related to the reduced interchain coupling. Here we are unable to make a reasonable fit for the $C_{\text{e}}(T)$ dependence because of limitation of the temperature range (only down to 0.43 $T_{\text{c}}$).

To examine whether the superconducting gap has nodes, we investigated the magnetic field dependence of the electronic specific heat, $C_{\text{e}}(T,H)$, which gives the information of low-energy excitations in the superconducting state. For a conventional fully gapped type-II superconductor, the excited states are basically the 'normal-state quasiparticles' in the vortex cores. This generates an additional Sommerfeld coefficient of $\gamma(H) \approx \frac{H}{H_{c2}(0)}\gamma_{\text{n}}$ at zero temperature in the superconducting mixed state.\cite{caroli} At zero field, the linear extrapolation of $C/T$ vs $T^2$ gives a 'residual' Sommerfeld coefficent of $\gamma_{0} = 5.3 $ mJ K$^{-2}$ mol$^{-1}$, which could be either due to the existence of small fraction ($\sim$ 10\%) of non-superconducting phase, or because of the existence of quasiparticles from impurity scattering in the case of nodal superconductivity. With increasing magnetic fields, $\gamma(H)$ increases faster than the full-gap prediction, especially at lower fields. Further, $[\gamma(H)-\gamma_0]$ basically obeys the Volovik relation (i.e., $\propto\sqrt{H}$),\cite{volovik} suggesting presence of nodes in the superconducting gap.

It is informative to compare the two the Cr$_3$As$_3$-chain based superconductors, K$_2$Cr$_3$As$_3$ and Rb$_2$Cr$_3$As$_3$, whose structural and physical parameters listed in Table ~\ref{parameter}. The 'anisotropic' expansion of the $a$ axis suggests a weaker interchain coupling in Rb$_2$Cr$_3$As$_3$, which could account for the broadening of superconducting transition and a reduced $\Delta C/(\gamma_{\text{n}} T_{\text{c}})$ value. Nevertheless, superconductivity emerges at a comparable temperature, and the variation of $T_{\text{c}}$ (decreased by 21\%) basically scales with the change in $\gamma_{\text{n}}$ (decreased by 22\%) which measures the DOS at $E_{\text{F}}$. If this is not an incidental phenomenon, one may expect higher $T_{\text{c}}$ for a larger $\gamma_{\text{n}}$ value. The estimated upper critical field exceeding the Pauli limit by a factor of two can be a hint for spin-triplet Cooper pairing. Besides, the existence of nodes in the superconducting order parameter, as suggested by the thermodynamic analysis, does not contradict with the spin-triplet scenario. Although much work needs to be done for the details of the superconducting order parameter, the fact that the only two existed members\cite{members} in the Q1D Cr-based family so far both show evidences of unconventional SC suggests that the unconventionality should be a common phenomenon in the Cr$_3$As$_3$-chain based system.

\begin{acknowledgments}
This work was supported by the National Basic Research Program of China
(Grant No. 2011CBA00103), the National Science Foundation of China (Grant No. 11190023), and the Fundamental Research Funds for the Central Universities of China.
\end{acknowledgments}

%References
\bibliography{233_Tang}

\begin{thebibliography}{27}%
\makeatletter
\providecommand \@ifxundefined [1]{%
 \@ifx{#1\undefined}
}%
\providecommand \@ifnum [1]{%
 \ifnum #1\expandafter \@firstoftwo
 \else \expandafter \@secondoftwo
 \fi
}%
\providecommand \@ifx [1]{%
 \ifx #1\expandafter \@firstoftwo
 \else \expandafter \@secondoftwo
 \fi
}%
\providecommand \natexlab [1]{#1}%
\providecommand \enquote  [1]{``#1''}%
\providecommand \bibnamefont  [1]{#1}%
\providecommand \bibfnamefont [1]{#1}%
\providecommand \citenamefont [1]{#1}%
\providecommand \href@noop [0]{\@secondoftwo}%
\providecommand \href [0]{\begingroup \@sanitize@url \@href}%
\providecommand \@href[1]{\@@startlink{#1}\@@href}%
\providecommand \@@href[1]{\endgroup#1\@@endlink}%
\providecommand \@sanitize@url [0]{\catcode `\\12\catcode `\$12\catcode
  `\&12\catcode `\#12\catcode `\^12\catcode `\_12\catcode `\%12\relax}%
\providecommand \@@startlink[1]{}%
\providecommand \@@endlink[0]{}%
\providecommand \url  [0]{\begingroup\@sanitize@url \@url }%
\providecommand \@url [1]{\endgroup\@href {#1}{\urlprefix }}%
\providecommand \urlprefix  [0]{URL }%
\providecommand \Eprint [0]{\href }%
\providecommand \doibase [0]{http://dx.doi.org/}%
\providecommand \selectlanguage [0]{\@gobble}%
\providecommand \bibinfo  [0]{\@secondoftwo}%
\providecommand \bibfield  [0]{\@secondoftwo}%
\providecommand \translation [1]{[#1]}%
\providecommand \BibitemOpen [0]{}%
\providecommand \bibitemStop [0]{}%
\providecommand \bibitemNoStop [0]{.\EOS\space}%
\providecommand \EOS [0]{\spacefactor3000\relax}%
\providecommand \BibitemShut  [1]{\csname bibitem#1\endcsname}%
\let\auto@bib@innerbib\@empty
%</preamble>
\bibitem [{\citenamefont {van Smaalen}(2005)}]{smaalen}%
  \BibitemOpen
  \bibfield  {author} {\bibinfo {author} {\bibfnamefont {S.}~\bibnamefont {van
  Smaalen}},\ }\href {\doibase 10.1107/s0108767304025437} {\bibfield  {journal}
  {\bibinfo  {journal} {Acta Crystallographica Section A}\ }\textbf {\bibinfo
  {volume} {61}},\ \bibinfo {pages} {51} (\bibinfo {year} {2005})}\BibitemShut
  {NoStop}%
\bibitem [{\citenamefont {Gruner}(1988)}]{gruner}%
  \BibitemOpen
  \bibfield  {author} {\bibinfo {author} {\bibfnamefont {G.}~\bibnamefont
  {Gruner}},\ }\href {\doibase 10.1103/RevModPhys.60.1129} {\bibfield
  {journal} {\bibinfo  {journal} {Reviews of Modern Physics}\ }\textbf
  {\bibinfo {volume} {60}},\ \bibinfo {pages} {1129} (\bibinfo {year}
  {1988})}\BibitemShut {NoStop}%
\bibitem [{\citenamefont {Voit}(1995)}]{voit}%
  \BibitemOpen
  \bibfield  {author} {\bibinfo {author} {\bibfnamefont {J.}~\bibnamefont
  {Voit}},\ }\href {\doibase 10.1088/0034-4885/58/9/002} {\bibfield  {journal}
  {\bibinfo  {journal} {Reports on Progress in Physics}\ }\textbf {\bibinfo
  {volume} {58}},\ \bibinfo {pages} {977} (\bibinfo {year} {1995})}\BibitemShut
  {NoStop}%
\bibitem [{\citenamefont {Jerome}\ \emph {et~al.}(1980)\citenamefont {Jerome},
  \citenamefont {Mazaud}, \citenamefont {Ribault},\ and\ \citenamefont
  {Bechgaard}}]{jerome}%
  \BibitemOpen
  \bibfield  {author} {\bibinfo {author} {\bibfnamefont {D.}~\bibnamefont
  {Jerome}}, \bibinfo {author} {\bibfnamefont {A.}~\bibnamefont {Mazaud}},
  \bibinfo {author} {\bibfnamefont {M.}~\bibnamefont {Ribault}}, \ and\
  \bibinfo {author} {\bibfnamefont {K.}~\bibnamefont {Bechgaard}},\ }\href {<Go
  to ISI>://WOS:A1980JG10700005} {\bibfield  {journal} {\bibinfo  {journal}
  {Journal De Physique Lettres}\ }\textbf {\bibinfo {volume} {41}},\ \bibinfo
  {pages} {L95} (\bibinfo {year} {1980})}\BibitemShut {NoStop}%
\bibitem [{\citenamefont {Zhang}\ and\ \citenamefont {De~Melo}(2007)}]{wzhang}%
  \BibitemOpen
  \bibfield  {author} {\bibinfo {author} {\bibfnamefont {W.}~\bibnamefont
  {Zhang}}\ and\ \bibinfo {author} {\bibfnamefont {C.~A. R.~S.}\ \bibnamefont
  {De~Melo}},\ }\href {\doibase 10.1080/00018730701403190} {\bibfield
  {journal} {\bibinfo  {journal} {Advances in Physics}\ }\textbf {\bibinfo
  {volume} {56}},\ \bibinfo {pages} {545} (\bibinfo {year} {2007})}\BibitemShut
  {NoStop}%
\bibitem [{\citenamefont {Greenblatt}\ \emph {et~al.}(1984)\citenamefont
  {Greenblatt}, \citenamefont {McCarroll}, \citenamefont {Neifeld},
  \citenamefont {Croft},\ and\ \citenamefont {Waszczak}}]{greenblatt}%
  \BibitemOpen
  \bibfield  {author} {\bibinfo {author} {\bibfnamefont {M.}~\bibnamefont
  {Greenblatt}}, \bibinfo {author} {\bibfnamefont {W.~H.}\ \bibnamefont
  {McCarroll}}, \bibinfo {author} {\bibfnamefont {R.}~\bibnamefont {Neifeld}},
  \bibinfo {author} {\bibfnamefont {M.}~\bibnamefont {Croft}}, \ and\ \bibinfo
  {author} {\bibfnamefont {J.~V.}\ \bibnamefont {Waszczak}},\ }\href {\doibase
  10.1016/0038-1098(84)90944-x} {\bibfield  {journal} {\bibinfo  {journal}
  {Solid State Communications}\ }\textbf {\bibinfo {volume} {51}},\ \bibinfo
  {pages} {671} (\bibinfo {year} {1984})}\BibitemShut {NoStop}%
\bibitem [{\citenamefont {Armici}\ \emph {et~al.}(1980)\citenamefont {Armici},
  \citenamefont {Decroux}, \citenamefont {Fischer}, \citenamefont {Potel},
  \citenamefont {Chevrel},\ and\ \citenamefont {Sergent}}]{armici}%
  \BibitemOpen
  \bibfield  {author} {\bibinfo {author} {\bibfnamefont {J.~C.}\ \bibnamefont
  {Armici}}, \bibinfo {author} {\bibfnamefont {M.}~\bibnamefont {Decroux}},
  \bibinfo {author} {\bibfnamefont {O.}~\bibnamefont {Fischer}}, \bibinfo
  {author} {\bibfnamefont {M.}~\bibnamefont {Potel}}, \bibinfo {author}
  {\bibfnamefont {R.}~\bibnamefont {Chevrel}}, \ and\ \bibinfo {author}
  {\bibfnamefont {M.}~\bibnamefont {Sergent}},\ }\href {\doibase
  10.1016/0038-1098(80)90734-6} {\bibfield  {journal} {\bibinfo  {journal}
  {Solid State Communications}\ }\textbf {\bibinfo {volume} {33}},\ \bibinfo
  {pages} {607} (\bibinfo {year} {1980})}\BibitemShut {NoStop}%
\bibitem [{\citenamefont {Denlinger}\ \emph {et~al.}(1999)\citenamefont
  {Denlinger}, \citenamefont {Gweon}, \citenamefont {Allen}, \citenamefont
  {Olson}, \citenamefont {Marcus}, \citenamefont {Schlenker},\ and\
  \citenamefont {Hsu}}]{denlinger}%
  \BibitemOpen
  \bibfield  {author} {\bibinfo {author} {\bibfnamefont {J.~D.}\ \bibnamefont
  {Denlinger}}, \bibinfo {author} {\bibfnamefont {G.~H.}\ \bibnamefont
  {Gweon}}, \bibinfo {author} {\bibfnamefont {J.~W.}\ \bibnamefont {Allen}},
  \bibinfo {author} {\bibfnamefont {C.~G.}\ \bibnamefont {Olson}}, \bibinfo
  {author} {\bibfnamefont {J.}~\bibnamefont {Marcus}}, \bibinfo {author}
  {\bibfnamefont {C.}~\bibnamefont {Schlenker}}, \ and\ \bibinfo {author}
  {\bibfnamefont {L.~S.}\ \bibnamefont {Hsu}},\ }\href {\doibase
  10.1103/PhysRevLett.82.2540} {\bibfield  {journal} {\bibinfo  {journal}
  {Physical Review Letters}\ }\textbf {\bibinfo {volume} {82}},\ \bibinfo
  {pages} {2540} (\bibinfo {year} {1999})}\BibitemShut {NoStop}%
\bibitem [{\citenamefont {Xu}\ \emph {et~al.}(2009)\citenamefont {Xu},
  \citenamefont {Bangura}, \citenamefont {Analytis}, \citenamefont {Fletcher},
  \citenamefont {French}, \citenamefont {Shannon}, \citenamefont {He},
  \citenamefont {Zhang}, \citenamefont {Mandrus}, \citenamefont {Jin},\ and\
  \citenamefont {Hussey}}]{xxf}%
  \BibitemOpen
  \bibfield  {author} {\bibinfo {author} {\bibfnamefont {X.}~\bibnamefont
  {Xu}}, \bibinfo {author} {\bibfnamefont {A.~F.}\ \bibnamefont {Bangura}},
  \bibinfo {author} {\bibfnamefont {J.~G.}\ \bibnamefont {Analytis}}, \bibinfo
  {author} {\bibfnamefont {J.~D.}\ \bibnamefont {Fletcher}}, \bibinfo {author}
  {\bibfnamefont {M.~M.~J.}\ \bibnamefont {French}}, \bibinfo {author}
  {\bibfnamefont {N.}~\bibnamefont {Shannon}}, \bibinfo {author} {\bibfnamefont
  {J.}~\bibnamefont {He}}, \bibinfo {author} {\bibfnamefont {S.}~\bibnamefont
  {Zhang}}, \bibinfo {author} {\bibfnamefont {D.}~\bibnamefont {Mandrus}},
  \bibinfo {author} {\bibfnamefont {R.}~\bibnamefont {Jin}}, \ and\ \bibinfo
  {author} {\bibfnamefont {N.~E.}\ \bibnamefont {Hussey}},\ }\href {\doibase
  10.1103/PhysRevLett.102.206602} {\bibfield  {journal} {\bibinfo  {journal}
  {Physical Review Letters}\ }\textbf {\bibinfo {volume} {102}} (\bibinfo
  {year} {2009}),\ 10.1103/PhysRevLett.102.206602}\BibitemShut {NoStop}%
\bibitem [{\citenamefont {Mercure}\ \emph {et~al.}(2012)\citenamefont
  {Mercure}, \citenamefont {Bangura}, \citenamefont {Xu}, \citenamefont
  {Wakeham}, \citenamefont {Carrington}, \citenamefont {Walmsley},
  \citenamefont {Greenblatt},\ and\ \citenamefont {Hussey}}]{Hc2_LiMoO}%
  \BibitemOpen
  \bibfield  {author} {\bibinfo {author} {\bibfnamefont {J.~F.}\ \bibnamefont
  {Mercure}}, \bibinfo {author} {\bibfnamefont {A.~F.}\ \bibnamefont
  {Bangura}}, \bibinfo {author} {\bibfnamefont {X.}~\bibnamefont {Xu}},
  \bibinfo {author} {\bibfnamefont {N.}~\bibnamefont {Wakeham}}, \bibinfo
  {author} {\bibfnamefont {A.}~\bibnamefont {Carrington}}, \bibinfo {author}
  {\bibfnamefont {P.}~\bibnamefont {Walmsley}}, \bibinfo {author}
  {\bibfnamefont {M.}~\bibnamefont {Greenblatt}}, \ and\ \bibinfo {author}
  {\bibfnamefont {N.~E.}\ \bibnamefont {Hussey}},\ }\href {\doibase
  10.1103/PhysRevLett.108.187003} {\bibfield  {journal} {\bibinfo  {journal}
  {Physical Review Letters}\ }\textbf {\bibinfo {volume} {108}} (\bibinfo
  {year} {2012}),\ 10.1103/PhysRevLett.108.187003}\BibitemShut {NoStop}%
\bibitem [{\citenamefont {Lebed}\ and\ \citenamefont {Sepper}(2013)}]{lebed}%
  \BibitemOpen
  \bibfield  {author} {\bibinfo {author} {\bibfnamefont {A.~G.}\ \bibnamefont
  {Lebed}}\ and\ \bibinfo {author} {\bibfnamefont {O.}~\bibnamefont {Sepper}},\
  }\href {\doibase 10.1103/PhysRevB.87.100511} {\bibfield  {journal} {\bibinfo
  {journal} {Physical Review B}\ }\textbf {\bibinfo {volume} {87}} (\bibinfo
  {year} {2013}),\ 10.1103/PhysRevB.87.100511}\BibitemShut {NoStop}%
\bibitem [{\citenamefont {Bao}\ \emph {et~al.}(2014)\citenamefont {Bao},
  \citenamefont {Liu}, \citenamefont {Ma}, \citenamefont {Meng}, \citenamefont
  {Tang}, \citenamefont {Sun}, \citenamefont {Zhai}, \citenamefont {Jiang},
  \citenamefont {Xu},\ and\ \citenamefont {Cao}}]{bao}%
  \BibitemOpen
  \bibfield  {author} {\bibinfo {author} {\bibfnamefont {J.~K.}\ \bibnamefont
  {Bao}}, \bibinfo {author} {\bibfnamefont {J.~Y.}\ \bibnamefont {Liu}},
  \bibinfo {author} {\bibfnamefont {C.~W.}\ \bibnamefont {Ma}}, \bibinfo
  {author} {\bibfnamefont {Z.~H.}\ \bibnamefont {Meng}}, \bibinfo {author}
  {\bibfnamefont {Z.~T.}\ \bibnamefont {Tang}}, \bibinfo {author}
  {\bibfnamefont {Y.~L.}\ \bibnamefont {Sun}}, \bibinfo {author} {\bibfnamefont
  {H.~F.}\ \bibnamefont {Zhai}}, \bibinfo {author} {\bibfnamefont {C.~M.}\
  \bibnamefont {Jiang}, \bibfnamefont {H.~Feng}}, \bibinfo {author}
  {\bibfnamefont {Z.~A.}\ \bibnamefont {Xu}}, \ and\ \bibinfo {author}
  {\bibfnamefont {G.~H.}\ \bibnamefont {Cao}},\ }\href@noop {} {\bibfield
  {journal} {\bibinfo  {journal} {arXiv}\ ,\ \bibinfo {pages} {1412.0067}}
  (\bibinfo {year} {2014})}\BibitemShut {NoStop}%
\bibitem [{\citenamefont {Jiang}\ \emph {et~al.}(2014)\citenamefont {Jiang},
  \citenamefont {Cao},\ and\ \citenamefont {Cao}}]{jh}%
  \BibitemOpen
  \bibfield  {author} {\bibinfo {author} {\bibfnamefont {H.}~\bibnamefont
  {Jiang}}, \bibinfo {author} {\bibfnamefont {G.~H.}\ \bibnamefont {Cao}}, \
  and\ \bibinfo {author} {\bibfnamefont {C.}~\bibnamefont {Cao}},\ }\href@noop
  {} {\bibfield  {journal} {\bibinfo  {journal} {arXiv}\ ,\ \bibinfo {pages}
  {1412.1309}} (\bibinfo {year} {2014})}\BibitemShut {NoStop}%
\bibitem [{\citenamefont {Petrovic}\ \emph {et~al.}(2010)\citenamefont
  {Petrovic}, \citenamefont {Lortz}, \citenamefont {Santi}, \citenamefont
  {Decroux}, \citenamefont {Monnard}, \citenamefont {Fischer}, \citenamefont
  {Boeri}, \citenamefont {Andersen}, \citenamefont {Kortus}, \citenamefont
  {Salloum}, \citenamefont {Gougeon},\ and\ \citenamefont {Potel}}]{petrovic}%
  \BibitemOpen
  \bibfield  {author} {\bibinfo {author} {\bibfnamefont {A.~P.}\ \bibnamefont
  {Petrovic}}, \bibinfo {author} {\bibfnamefont {R.}~\bibnamefont {Lortz}},
  \bibinfo {author} {\bibfnamefont {G.}~\bibnamefont {Santi}}, \bibinfo
  {author} {\bibfnamefont {M.}~\bibnamefont {Decroux}}, \bibinfo {author}
  {\bibfnamefont {H.}~\bibnamefont {Monnard}}, \bibinfo {author} {\bibfnamefont
  {O.}~\bibnamefont {Fischer}}, \bibinfo {author} {\bibfnamefont
  {L.}~\bibnamefont {Boeri}}, \bibinfo {author} {\bibfnamefont {O.~K.}\
  \bibnamefont {Andersen}}, \bibinfo {author} {\bibfnamefont {J.}~\bibnamefont
  {Kortus}}, \bibinfo {author} {\bibfnamefont {D.}~\bibnamefont {Salloum}},
  \bibinfo {author} {\bibfnamefont {P.}~\bibnamefont {Gougeon}}, \ and\
  \bibinfo {author} {\bibfnamefont {M.}~\bibnamefont {Potel}},\ }\href
  {\doibase 10.1103/PhysRevB.82.235128} {\bibfield  {journal} {\bibinfo
  {journal} {Physical Review B}\ }\textbf {\bibinfo {volume} {82}} (\bibinfo
  {year} {2010}),\ 10.1103/PhysRevB.82.235128}\BibitemShut {NoStop}%
\bibitem [{\citenamefont {Izumi}\ and\ \citenamefont
  {Momma}(2007)}]{rietan-fp}%
  \BibitemOpen
  \bibfield  {author} {\bibinfo {author} {\bibfnamefont {F.}~\bibnamefont
  {Izumi}}\ and\ \bibinfo {author} {\bibfnamefont {K.}~\bibnamefont {Momma}},\
  }\href@noop {} {\bibfield  {journal} {\bibinfo  {journal} {Solid State
  Phenomena}\ }\textbf {\bibinfo {volume} {130}},\ \bibinfo {pages} {15}
  (\bibinfo {year} {2007})}\BibitemShut {NoStop}%
\bibitem [{pol()}]{polycrystal}%
  \BibitemOpen
  \href@noop {} {}\bibinfo {note} {Due to the anisotropy in resistivity, the
  as-measured $\rho(T)$ using polycrystalline samples should be a "combined"
  result of $\rho_{\|}(T)$ (resistivity along the chain direction) and
  $\rho_{\bot}(T)$, as well as the contribution from the grain boundaries.
  Thus, the intrinsic $\rho_{\|}(T)$ behavior needs to be examined by using the
  single-crystal sample without any deterioration.}\BibitemShut {Stop}%
\bibitem [{NMR()}]{NMR}%
  \BibitemOpen
  \href@noop {} {}\bibinfo {note} {Very recently, the $^{75}$As nuclear
  spin-lattice relaxation rate data show the evidence of Tomonaga-Luttinger
  liquid state in K$_2$Cr$_3$As$_3$ [H. Z. Zhi et al., arXiv: 1501.00713
  (2015)].}\BibitemShut {Stop}%
\bibitem [{\citenamefont {Ogata}\ and\ \citenamefont
  {Anderson}(1993)}]{ogata-anderson}%
  \BibitemOpen
  \bibfield  {author} {\bibinfo {author} {\bibfnamefont {M.}~\bibnamefont
  {Ogata}}\ and\ \bibinfo {author} {\bibfnamefont {P.~W.}\ \bibnamefont
  {Anderson}},\ }\href {\doibase 10.1103/PhysRevLett.70.3087} {\bibfield
  {journal} {\bibinfo  {journal} {Physical Review Letters}\ }\textbf {\bibinfo
  {volume} {70}},\ \bibinfo {pages} {3087} (\bibinfo {year}
  {1993})}\BibitemShut {NoStop}%
\bibitem [{\citenamefont {Ginzburg}\ and\ \citenamefont {Landau}(1950)}]{GL}%
  \BibitemOpen
  \bibfield  {author} {\bibinfo {author} {\bibfnamefont {V.~L.}\ \bibnamefont
  {Ginzburg}}\ and\ \bibinfo {author} {\bibfnamefont {L.~D.~Z.}\ \bibnamefont
  {Landau}},\ }\href@noop {} {\bibfield  {journal} {\bibinfo  {journal} {Eksp.
  Teor. Fiz.}\ }\textbf {\bibinfo {volume} {20}},\ \bibinfo {pages} {1064}
  (\bibinfo {year} {1950})}\BibitemShut {NoStop}%
\bibitem [{\citenamefont {Werthame}\ \emph {et~al.}(1966)\citenamefont
  {Werthame}, \citenamefont {Helfand},\ and\ \citenamefont {Hohenber}}]{WHH}%
  \BibitemOpen
  \bibfield  {author} {\bibinfo {author} {\bibfnamefont {N.~R.}\ \bibnamefont
  {Werthame}}, \bibinfo {author} {\bibfnamefont {E.}~\bibnamefont {Helfand}}, \
  and\ \bibinfo {author} {\bibfnamefont {P.~C.}\ \bibnamefont {Hohenber}},\
  }\href {\doibase 10.1103/PhysRev.147.295} {\bibfield  {journal} {\bibinfo
  {journal} {Physical Review}\ }\textbf {\bibinfo {volume} {147}},\ \bibinfo
  {pages} {295} (\bibinfo {year} {1966})}\BibitemShut {NoStop}%
\bibitem [{\citenamefont {Hake}(1967)}]{hake}%
  \BibitemOpen
  \bibfield  {author} {\bibinfo {author} {\bibfnamefont {R.~R.}\ \bibnamefont
  {Hake}},\ }\href {\doibase http://dx.doi.org/10.1063/1.1754905} {\bibfield
  {journal} {\bibinfo  {journal} {Applied Physics Letters}\ }\textbf {\bibinfo
  {volume} {10}},\ \bibinfo {pages} {189} (\bibinfo {year} {1967})}\BibitemShut
  {NoStop}%
\bibitem [{\citenamefont {Clogston}(1962)}]{clogston}%
  \BibitemOpen
  \bibfield  {author} {\bibinfo {author} {\bibfnamefont {A.~M.}\ \bibnamefont
  {Clogston}},\ }\href {\doibase 10.1103/PhysRevLett.9.266} {\bibfield
  {journal} {\bibinfo  {journal} {Physical Review Letters}\ }\textbf {\bibinfo
  {volume} {9}},\ \bibinfo {pages} {266} (\bibinfo {year} {1962})}\BibitemShut
  {NoStop}%
\bibitem [{\citenamefont {Chandrasekhar}(1962)}]{chandrasekhar}%
  \BibitemOpen
  \bibfield  {author} {\bibinfo {author} {\bibfnamefont {B.~S.}\ \bibnamefont
  {Chandrasekhar}},\ }\href {\doibase 10.1063/1.1777362} {\bibfield  {journal}
  {\bibinfo  {journal} {Applied Physics Letters}\ }\textbf {\bibinfo {volume}
  {1}},\ \bibinfo {pages} {7} (\bibinfo {year} {1962})}\BibitemShut {NoStop}%
\bibitem [{Ce()}]{Ce}%
  \BibitemOpen
  \href@noop {} {}\bibinfo {note} {In most cases, the Schottky-anomaly
  contribution is negligibly small above 2 K.}\BibitemShut {Stop}%
\bibitem [{\citenamefont {Caroli}\ \emph {et~al.}(1964)\citenamefont {Caroli},
  \citenamefont {Degennes},\ and\ \citenamefont {Matricon}}]{caroli}%
  \BibitemOpen
  \bibfield  {author} {\bibinfo {author} {\bibfnamefont {C.}~\bibnamefont
  {Caroli}}, \bibinfo {author} {\bibfnamefont {P.~G.}\ \bibnamefont
  {Degennes}}, \ and\ \bibinfo {author} {\bibfnamefont {J.}~\bibnamefont
  {Matricon}},\ }\href {\doibase 10.1016/0031-9163(64)90375-0} {\bibfield
  {journal} {\bibinfo  {journal} {Physics Letters}\ }\textbf {\bibinfo {volume}
  {9}},\ \bibinfo {pages} {307} (\bibinfo {year} {1964})}\BibitemShut {NoStop}%
\bibitem [{\citenamefont {Volovik}(1993)}]{volovik}%
  \BibitemOpen
  \bibfield  {author} {\bibinfo {author} {\bibfnamefont {G.~E.}\ \bibnamefont
  {Volovik}},\ }\href {<Go to ISI>://WOS:A1993MF77700013} {\bibfield  {journal}
  {\bibinfo  {journal} {JETP Letters}\ }\textbf {\bibinfo {volume} {58}},\
  \bibinfo {pages} {469} (\bibinfo {year} {1993})}\BibitemShut {NoStop}%
\bibitem [{mem()}]{members}%
  \BibitemOpen
  \href@noop {} {}\bibinfo {note} {The third member of the Cr-based family,
  Cs$_2$Cr$_3$As$_3$, was synthesized very recently [Z. T. Tang et al., arXiv:
  1501.02065 (2015)]. It shows possibly unconventional superconductivity at 2.2
  K.}\BibitemShut {Stop}%
\end{thebibliography}%

\end{document}